\documentstyle[12pt]{article}
\begin{document}
\title{\bf Inelasticity in hadron-nucleus collisions from emulsion
chamber studies }
\author{G.Wilk$^{1}$\thanks{e-mail: wilk@fuw.edu.pl} and Z.W\l
odarczyk$^{2}$ \thanks{e-mail: wlod@pu.kielce.pl}\\[2ex] 
$^{1}${\it The Andrzej So\l tan Institute for Nuclear Studies}\\
{\it Ho\.za 69; 00-689 Warsaw, Poland}\\
$^{2}${\it Institute of Physics, Pedagogical University}\\
{\it Konopnickiej 15, 25-405 Kielce, Poland}}
\date{\today}
\maketitle

\begin{abstract}
The inelasticity of hadron-carbon nucleus collisions in the energy
region exceeding $100$ TeV is estimated from the carbon-emulsion
chamber data at Pamirs to be $\langle K_C\rangle = 0.65\pm 0.08$.
When combined with the recently presented data on hadron-lead nucleus
collisions taken at the same energy range it results in the $K\sim
A^{0.086}$ mass number dependence of inelasticity. The evaluated
partial inelasticity for secondary ($\nu > 1$) interactions, $K_{\nu
>1} \simeq 0.2$, suggests that the second and higher interactions of
the excited hadron inside the nucleus proceed with only slight energy
losses. \\

\end{abstract}

\newpage
\section{Introduction}

The inelasticity of hadronic reactions, understood as the fraction of
the incident beam energy not carried off by fragments of the
projectile, is (next to the inelastic cross section) the most
significant variable for all cosmic ray experiments involved in
cascade developments \cite{SWWW,FGS}. The low energy data (in the
$100 - 200$ GeV range) show that the stopping power of nuclei is
rather low \cite{TT,FGS}. At higher energies there is no accelerator
data for inelasticity \cite{ZIK} and only rough indications from
cosmic ray experiments are available \cite{SWWW,FGS}. Recently
\cite{B} the inelasticity in hadron-lead collisions was estimated in
the energy region exceeding $100$ TeV. In the present paper we
discuss hadron-carbon nucleus collisions observed by carbon emulsion
chamber, which are exposed to cosmic rays at the Pamirs. In the next
Section we present the experimental method used (which is similar to
that used in \cite{B} and more starightforward than the one explored
in \cite{TT}). Section 3 contains our results, which, when combined
with those of \cite{B}, allow us to deduce the mass
number dependence of inelasticity directly from experimental data. In
Section 4 we discuss the (model dependent) notion of partial
inelasticity providing the information on the character of secondary
interactions in the nuclei (albeit in a model dependent way). Last
Section summarizes and concludes our presentation.\\

\section{Experimental method - repeated registration of cascades}

In the Pamir experiments, among others, a multi-layer $X$-ray film
emulsion chambers (EC) with large area two-carbon-generators (the
so-called hadronic ($H$) blocks) have been exposed \cite{W}. The
carbon chamber designed to observe hadrons consisted of:
$\Gamma$-block of $6$ cm $Pb$ (corresponding to 
$0.35 \lambda$  and $10.5$ c.u.) and two
$H$-blocks of carbon layer of $60$ cm thickness each ($66 {\rm
g}/{\rm cm}^2, 0.9\lambda, 2.5$ c.u.) followed by $5$ cm of
lead-emulsion sandwiches, cf. Fig. 1. In EC (which is a shallow
calorimeter) only the energy transfered to the electromagnetic
component is measured: 
\begin{equation}
E_h ^{\gamma}\, =\, K_{\gamma} \cdot E_h \label{eq:kg},
\end{equation}
(here coefficient $K_{\gamma}$ denotes the respective
electromagnetic part of the inelasticity)
and in the hadronic block a given nuclear-electromagnetic cascade
(NEC) produces spots with optical density $D$ on $X$-ray film.
General methodical problem of hadronic block measurements of how
to obtain the energy of the incoming hadron, $E_h$, from data on the
optical densities $D$, i.e., the transition: $D\longrightarrow
E_h^{\gamma} \longrightarrow E_h$, was examined in \cite{TW}. \\  

This specific structure of the carbon-emulsion chamber allows for a
relatively straightforward estimation of the total inelasticity for 
hadron-carbon nucleus interactions. Although such a possibility was
pointed out already in Ref.\cite{NWP} it was not utilised so far. We
shall use it now to estimate inelasticity for hadron-carbon interactions
at energies exceeding $100$ TeV. The proposed method is connected with
the repeated registration of the same NEC in the two subsequent
hadronic blocks.  
If $N_1$ denotes the number of cascades registered in the first
hadronic block (each cascade with visible energy greater than
some threshold energy ($E_h^{\gamma})_1$) and $N_2$ denotes the number
of cascades repeatedly registered also in the second hadronic block
(each cascade with visible energy above the threshold
($E_h^{\gamma})_2$), then it turns out that the quantity
\begin{equation}
\eta \, =\, \frac{N_2}{N_1} \label{eq:eta}
\end{equation}
is sensitive to the total (mean) inelasticity $\langle K\rangle$.
Similarly, for each event where NEC develops both in the upper and
lower $H$-blocks depositing there energies $E_1$ and $E_2$,
respectively, the ratio 
\begin{equation}
\epsilon\, =\, \frac{E_2}{E_1} \label{eq:epsilon}
\end{equation}
also depends on $\langle K\rangle$.
The weak dependence of both quantities on the methodical errors
(which to large extend cancel in the ratio) and ease with which the
experimental data may by obtained render this method very useful and
promising for possible future applications. \\  

To illustrate sensitivity of both quantities, $\epsilon$ and $\eta$,
on the inelasticity let us first consider simplified case of 
monochromatic beam of nucleons of energy $E_0$ entering our EC and
let us neglect for a moment the NEC in the target. In this case for
each event we have:
\begin{equation}
\epsilon\, =\, \frac{E_2}{E_1}\, =\, \frac{\left(1 - \langle
K\rangle \right) \cdot E_0\cdot \langle K_{\gamma}\rangle}{\langle
K_{\gamma}\rangle\cdot E_0}\, =\, 1\, -\, \langle K\rangle ,
\label{eq:epsilons} 
\end{equation}
where $\langle K\rangle$ is the (mean) total inelasticity. Notice
that $\langle K_{\gamma}\rangle$ from the eq.(\ref{eq:kg}) drops out
from the ratio $\epsilon$. Similarly, the relative number $\eta$ of
hadrons repeatedly registered in the two subsequent $H$-blocks of
thickness $x/\lambda$ each is
\begin{equation}
\eta\, =\, \frac{N_2}{N_1}\, =\, 
        \frac{N_1 \left(1 - e^{- x/\lambda}\right)}{N_1}\cdot 
                   \Phi\left(\langle K\rangle\right)\, =\,
                   \left( 1\, -\, e^{- x/\lambda}\right)\cdot 
                   \Phi\left(\langle K\rangle\right), \label{eq:etas}
\end{equation}
where $\Phi = \int^1_{K_{min}} \varphi(K)\, dK$ accounts for the energy
thresholds $E^{th}_1$ and $E^{th}_2$, which lead to the fact that
from the inelasticity distribution $\varphi(K)$ only the inelasticities 
$ K\, >\,  K_{min} = 
         E^{th}_2/\left(1 - \langle K\rangle\right) E^{th}_1$
are observed. In the case of $\varphi(K) = {\rm const}$ one gets 
$\Phi(\langle K\rangle)\, =\, 1\, -\, K_{min}$.\\

However, in true event one has to account for the following facts:
\begin{itemize}
\item[$(i)$] The incoming cosmic ray flux is not monochromatic but has
typical energy spectrum $N(E_0) \sim E^{- \gamma}_0$ and all energies
should be considered. In the region of interest to us (i.e., at
the mountain altitudes and energy region where 
data were collected) $\gamma \simeq 3$ \cite{Geiser,W2,FOOT1}
\item[$(ii)$] Cosmic ray flux at mountain altitudes considered here
contains not only nucleons but also mesons produced
in previous cascading processes in the atmosphere \cite{FOOT2}.
\item[$(iii)$] In reality EC do not register individual hadrons but
rather NEC developed by them. In Fig. 1 the incoming hadron
originates in the upper $H$-block NEC, which then develops further.
Its electromagnetic component is registered as visible energies $E_1$
and $E_2$ (cf. eq.(\ref{eq:kg})) released in the upper and lower
$H$-blocks, respectively \cite{FOOT3}. Each cascade is therefore
recorded as single event with visible energies $E_1$ and $E_2$.
\end{itemize}
To account for these points one therefore has to resort to the Monte
Carlo simulation calculations.

\section{Inelasticity in hadron-carbon nucleus interactions}

The experimental data collected from $110 {\rm m}^2$ carbon EC
contain $N_1 = 70$ cascades with energies $E_1 > 30$ TeV 
among which $N_2 = 24$ cascades have energies $E_2 > 2$ TeV). They
give the value of $\eta = 0.27\pm 0.06$ (at energy threshold $E_2 >
4$ TeV, being free from the detection bias) and the energy ratio
$\epsilon = 0.24\pm 0.07$. These data have been then recalculated by
using the simulated $D(E_h^{\gamma})$ dependence \cite{TW}. The
repeated registrations of cascades has been simulated by the  
standard SHOWERSIM Monte-Carlo event generator \cite{WROT}.
Primary hadrons (assumed to consist of $75\%$ nucleons and $25\%$ pions
\cite{W1,FOOT2}) were sampled from the power spectrum representing
distribution of the initial energy with a differential slope equal to
$\gamma = 3$ \cite{Geiser,W2}. In each cascade gamma quanta and
electrons above $0.01$ TeV, reaching the detection level within the
radius of $5$ mm, were recorded and the corresponding optical
densities were calculated within the radii utilized in the
experiment. Only cascades with the energies above $E_1 = 30$ TeV and
$E_2 = 2$ TeV were selected.\\ 

The ratio $\eta$ of the number of cascades repeatedly registered in
two hadronic blocks and the number of all cascades registered in the
first hadronic block is presented in Fig. 2 for different total
inelasticities: $\langle K\rangle = 0.5, 0.65$ and $0.80$. Note that
the ratio $\eta$ is more sensitive to the mean value of inelasticity
$\langle K\rangle$ than the energy ratio $\epsilon$, shown for
illustration in Fig. 3. In Fig. 4 we show the $\chi ^2$ per degree of
freedom obtained for $\eta$ fits plotted as a function of the assumed
inelasticity $K$. The comparison of experimental data with simulated
dependences indicates that $\langle K_C\rangle = 0.65 \pm 0.08$ for
hadron-carbon nucleus collisions at the hadron energies of above
$\sim 100$ TeV is most probably choice for the mean value of
inelasticity at this energy for hadron-carbon collisions. This
consist the main result of our work.\\ 

Recently analysis of similar succesive hadron interactions registered
in other emulsion chambers exposed at Pamirs, in the so called
thick-lead-emulsion chambers ($60$ cm Pb or $3.2$ mean free paths
$\lambda$ of inelastic collision of nucleon) have been reported
\cite{B}. The corresponding inelasticity distribution of hadron-lead
collisions in the energy region exceeding $100$ TeV was estimated by
using distribution of the energy ratio $z = E_1 / \sum E_i$ obtained
from $74$ events of hadron interactions. The resulting average value
of the inelasticity is $\langle K_{Pb}\rangle = 0.83 \pm 0.17$.
Comparing now this result with our estimation of inelasticity for
hadron-carbon nucleus results in the following mass number dependence
of inelasticity: $K \sim A^{0.086}$.\\

\section{Partial inelasticity $K_{\nu}$}

Following the work of Ref. \cite{FGS} we shall now consider for
hadron-nucleus collision the so called partial inelasticities
$K_{\nu}$. This is model dependent quantity and in the framework of
Glauber multiple scattering formalism \cite{G} it is defined in the
following way: 
\begin{equation}
\langle 1 - K\rangle \, =\, \sum_{\nu = 1}\,  P_{\nu} 
                          \prod_{i=1}^{\nu}\, 
                          \langle 1 - K_i\rangle 
\end{equation}
where $P_{\nu}$ is the probability for encountering exactly $\nu$
wounded nucleons in a target of mass $A$ and $\langle 1 - K_i
\rangle$  is the mean elasticity of the leading hadron in the
encounter with the $i^{th}$ wounded nucleon. We assume now that partial
inelasticity $K_1$ is determined by hadron-proton scattering and
shall treat the remaining partial inelasticities $K_{\nu >1} = K_2$
as one free parameter \cite{FGS} constrained by fitting the $h$-nucleus
data. The total elasticity can be now written as
\begin{equation}
\langle 1 - K\rangle \, =\, \left( 1 - K_1\right)\, \sum_{\nu = 1}
\langle 1 - K_2 \rangle ^{\nu - 1}\,  P_{\nu } \, .
\end{equation}
The ratio of elasticities in collisions on Pb and C targets,
\begin{equation}
 \kappa \, =\, \frac{\langle 1 - K_{Pb}\rangle}
 {\langle 1 - K_C\rangle} \, ,
\end{equation}
depends only on $K_2$ once the $P_{\nu}$ is known. Assuming now, for
simplicity, Poisson distribution for the number of repeated collisions,
\begin{equation}
P_{\nu} = \frac{\langle \nu - 1 \rangle^{\nu - 1}}{(\nu - 1)!}
          \exp(-<\nu - 1>) \qquad ({\rm for}~~\nu = 1,2,\dots ,)
\end{equation}
we obtain that
\begin{equation}
\kappa \, =\, \frac{
	  \exp \left( - \langle \nu_{Pb} - 1\rangle\, K_2\right)}
         {\exp \left( - \langle \nu_{C} - 1\rangle\, K_2\right)}
 \qquad {\rm or}\qquad 
 K_2\, =\, \frac{-\, \ln \kappa}{\langle \nu_{Pb}\rangle\, -\, 
                                \langle \nu_{C} \rangle} \, .            
\end{equation}
In Fig. 5 we show, for different mass number dependence of mean number of
wounded nucleons as provided by the exponent $\alpha$ : $\langle \nu
\rangle \sim A^{\alpha}$, the dependence of the partial inelasticity
$K_2$ on the power index $\alpha$ and for the value of $\kappa =0.5$
which is obtained from the comparison of data on Pb and C nuclei. The
value of $K_2$ for the expected mean number of wounded nucleons,
$\langle \nu \rangle = A \sigma_{h-p}/\sigma_{hA} \sim A^{1/3}$, is
therefore equal to $K_2 \simeq 0.2$. Notice that there is tacit
assumption made here behind this value of partial inelasticity $K_2$,
namely that the ultimate identity of the final state nucleon is
determined only once during the interaction with the nucleus (what in
\cite{FGS} corresponds to the value $\beta = 1$ for the parameter
specifyng the fraction of isospin preserving reactions).\\

Our estimation of $K_2$ at energies above $100$ TeV is consistent
with low energy data (cf. Ref. \cite{FGS}). Note that inequality
$K_{\nu > 1} < K_1$ is characteristic to all string-type interaction
models (cf. Quark-Gluon String model \cite{QGS} or Dual Parton Model
\cite{DPM}). On the other hand the SIBYLL model \cite{FGS,SIBYLL}
predict much smaller value of $K_2$ in the examined energy 
region. In DPM and QGS models, when only one target nucleon is
wounded, a constituent quark (di-quark) belonging to the projectile
hadron couples to a string that in turn connects to a di-quark
(quark) belonging to the wounded nucleon. In the case where there are
two or more wounded nucleons in the target, the additional nucleons
can couple only to the sea quarks of the projectile. In this way the
desired physics can be reproduced by the model. In particular, the
excited hadron, being off mass-shell, does not interact repeatedly as
a physical hadron inside the nucleus.\\

\section{Summary}

For hadron-carbon nucleus collisions in energy region exceeding
$100$ TeV the inelasticity is estimated to be equal to $\langle
K_C\rangle = 0.65\pm 0.08$. This value, when compared with the value
$\langle K_{Pb}\rangle = 0.83\pm 0.17$ obtained recently for
hadron-lead collisions, results in the mass number dependence of
inelasticity given by $K \sim A^{0.086}$. Essentially the same
$A$-dependence has been reported in \cite{TT} (the lower values of 
inelasticities obtained there can be attributed to the fact that in
our case we estimate total inelasticity whereas in \cite{TT}
inelasticity was estimated more indirectly from the production and
distribution of charged secondary particles only). The evaluated
partial inelasticity $K_{\nu >1} = 0.2$ leads to the (model
dependent) conclusion that the second and higher interactions of the
excited hadron inside the nucleus are relatively elastic. Our
estimation of $K_{\nu >1}$ at energies above $100$ TeV \cite{FOOT4}
is consistent with the low energy data ($\sim 100$ GeV) and coincides
with the string-type model predictions. \\

\newpage
\noindent
{\bf Figure captions}
\begin{itemize}
\item[Fig.1.] The scheme of the carbon emulsion chamber with
a typical nuclear-electromagnetic cascade (NEC).
The incoming hadron initiates NEC in which leading particle and
secondary hadrons (solid lines) interact repeatedly while the
electromagnetic compnent, i.e., $\gamma$ quanta from $\pi^0$ decays
(broken lines), are recorded as total energies $E_1$ and $E_2$
deposited in the two lead-emulsion sandwiches following,
respectively, upper and lower $H$-blocks. Notice that in 
reality transverse dimensions of NEC are very small (of the order of
$100~\mu$m) and particles are not separated experimentally.

\item[Fig.2.] Dependence of $\eta = N_2/N_1$ on the energy threshold
$E_2^{th}$ in the second hadronic block for: $\langle K\rangle = 0.65$
(solid line), $\langle K\rangle =0.50$ (dotted line) and 
for $\langle K\rangle = 0.80$ (black dots), compared with the 
experimental data. 

\item[Fig.3.] Dependence of $\epsilon = E_2/E_1$ on the thickness
$H/\lambda$ of carbon target (the plotted curves correspond to
different $\langle K\rangle$ as in Fig. 2). The experimental point at
$H/\lambda = 1.1$ corresponds to our specific carbon emulsion chamber
(with inclusion of the averaging over zenith angle distribution of
incoming hadron which shifts the value $H/\lambda = 0.9$ to $1.1$).

\item[Fig.4.] The quality of $\eta - E_2^{th}$ fit shown by the
$\chi^2$ per degree of freedom (def) ploted versus the mean  
inelasticity of hadron-carbon nucleus collisions. 

\item[Fig.5.] The dependence of partial inelasticity $K_2$ on the
power index $\alpha$ (in the formula $<\nu>\sim A^{\alpha}$) 
for the experimental value of $\kappa = 0.5$.
\end{itemize}
\newpage
\centerline{Figure 1}
\begin{figure}[h]
\setlength{\unitlength}{1cm}
\begin{picture}(15.,10.)
\includegraphics{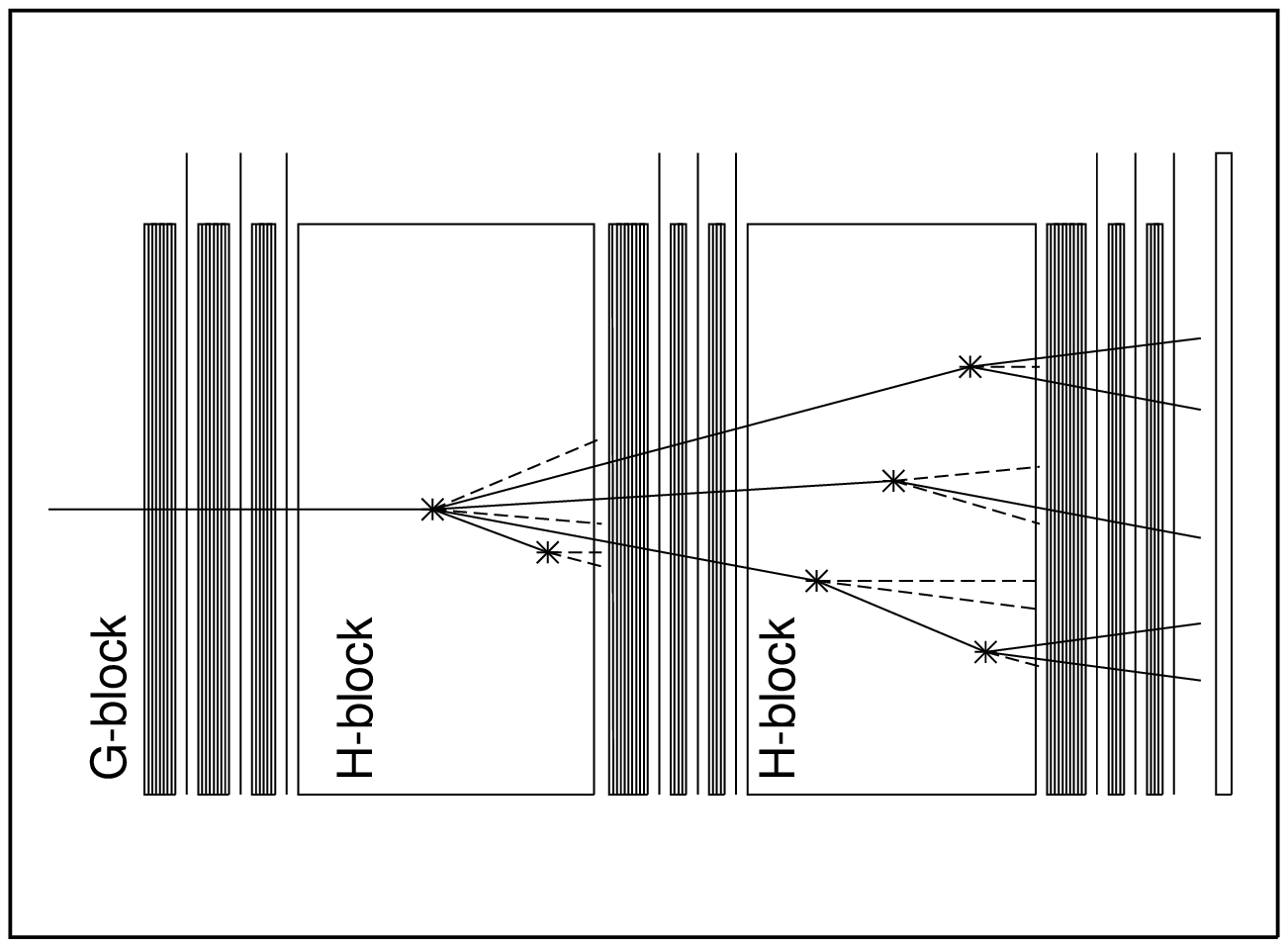}
\end{picture}
\end{figure}
\newpage
\centerline{Figure 2}
\begin{figure}[h]
\setlength{\unitlength}{1cm}
\begin{picture}(15.,10.)
\includegraphics{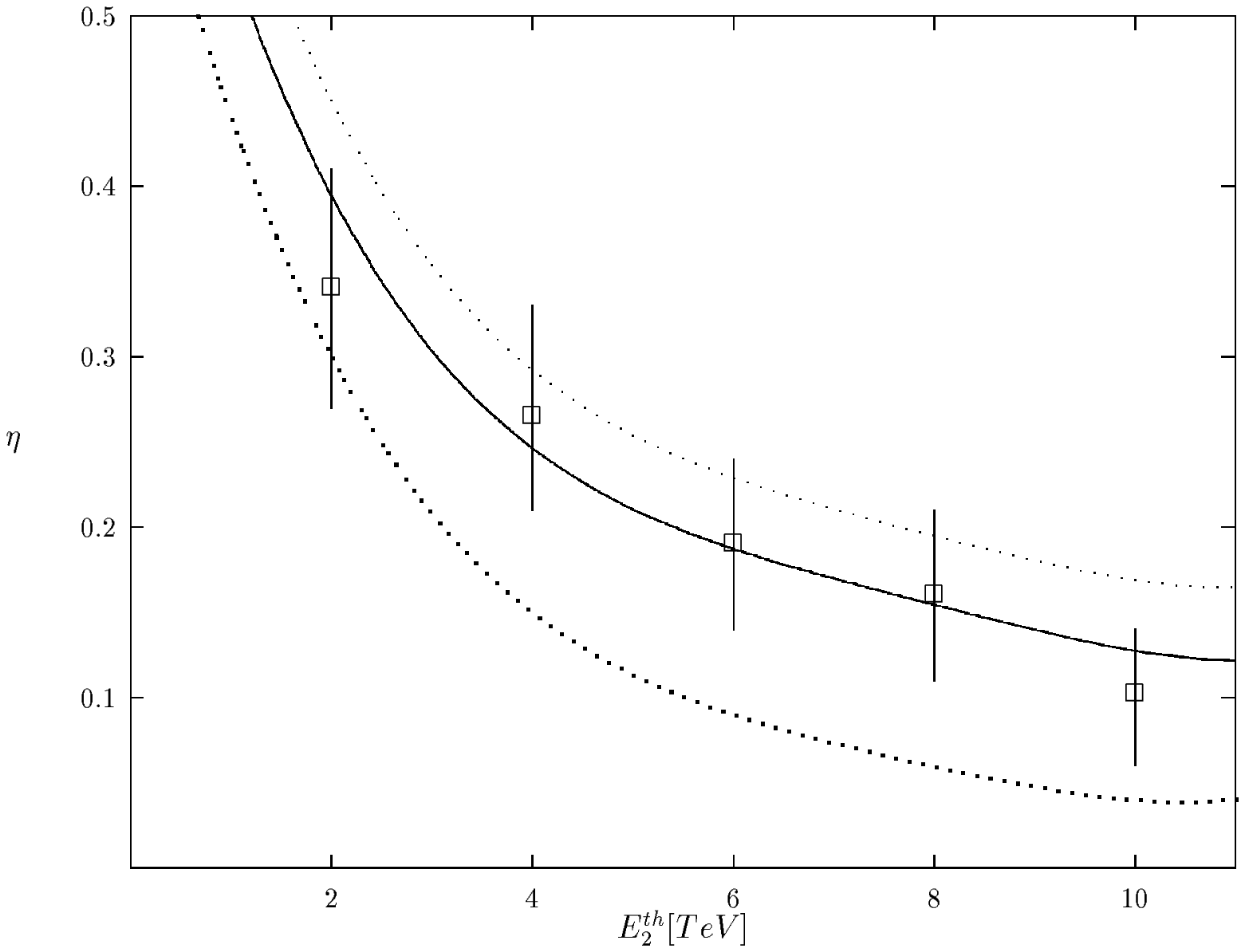}
\end{picture}
\end{figure}
\newpage
\centerline{Figure 3}
\begin{figure}[h]
\setlength{\unitlength}{1cm}
\begin{picture}(15.,10.)
\includegraphics{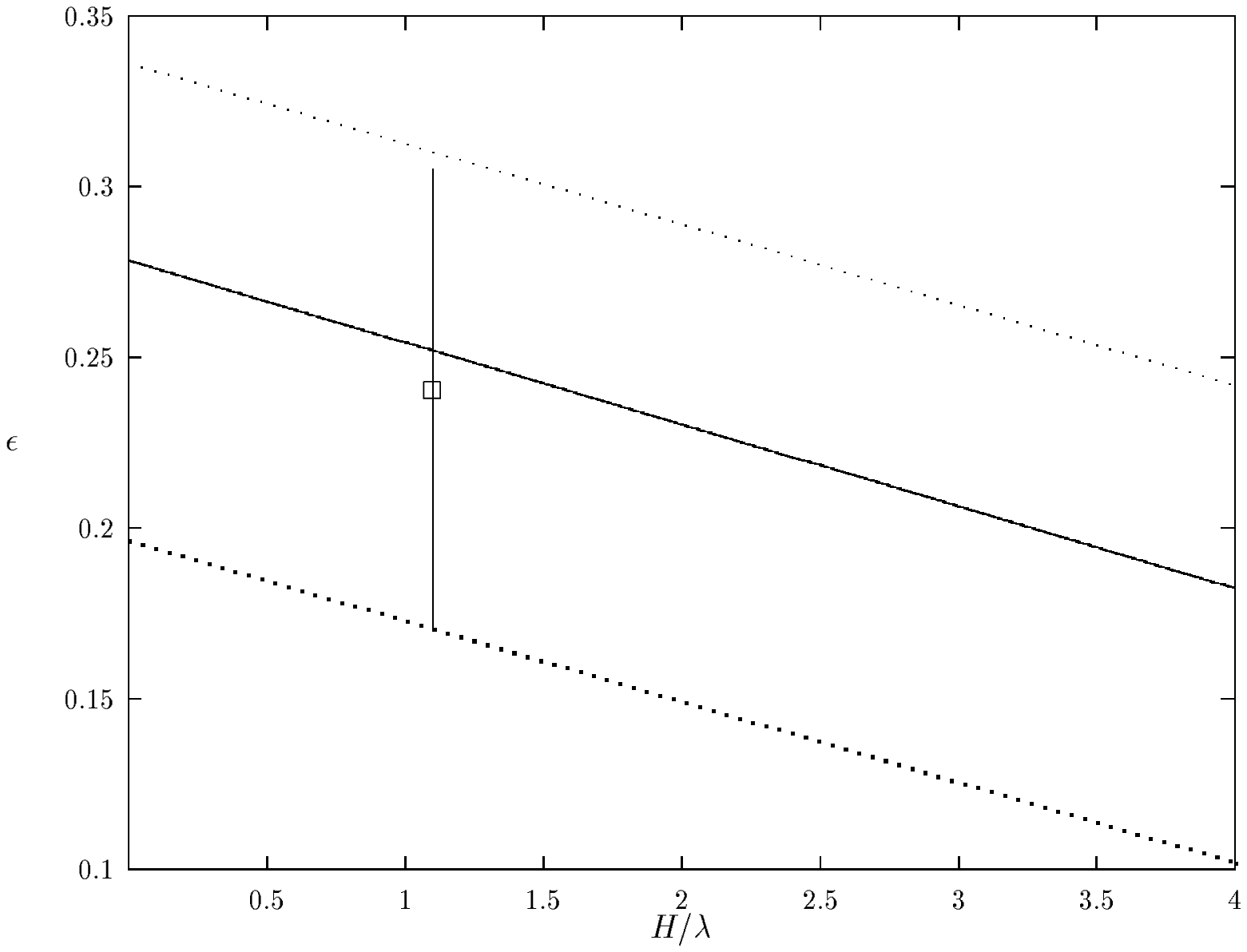}
\end{picture}
\end{figure}
\newpage
\centerline{Figure 4}
\begin{figure}[h]
\setlength{\unitlength}{1cm}
\begin{picture}(15.,10.)
\includegraphics{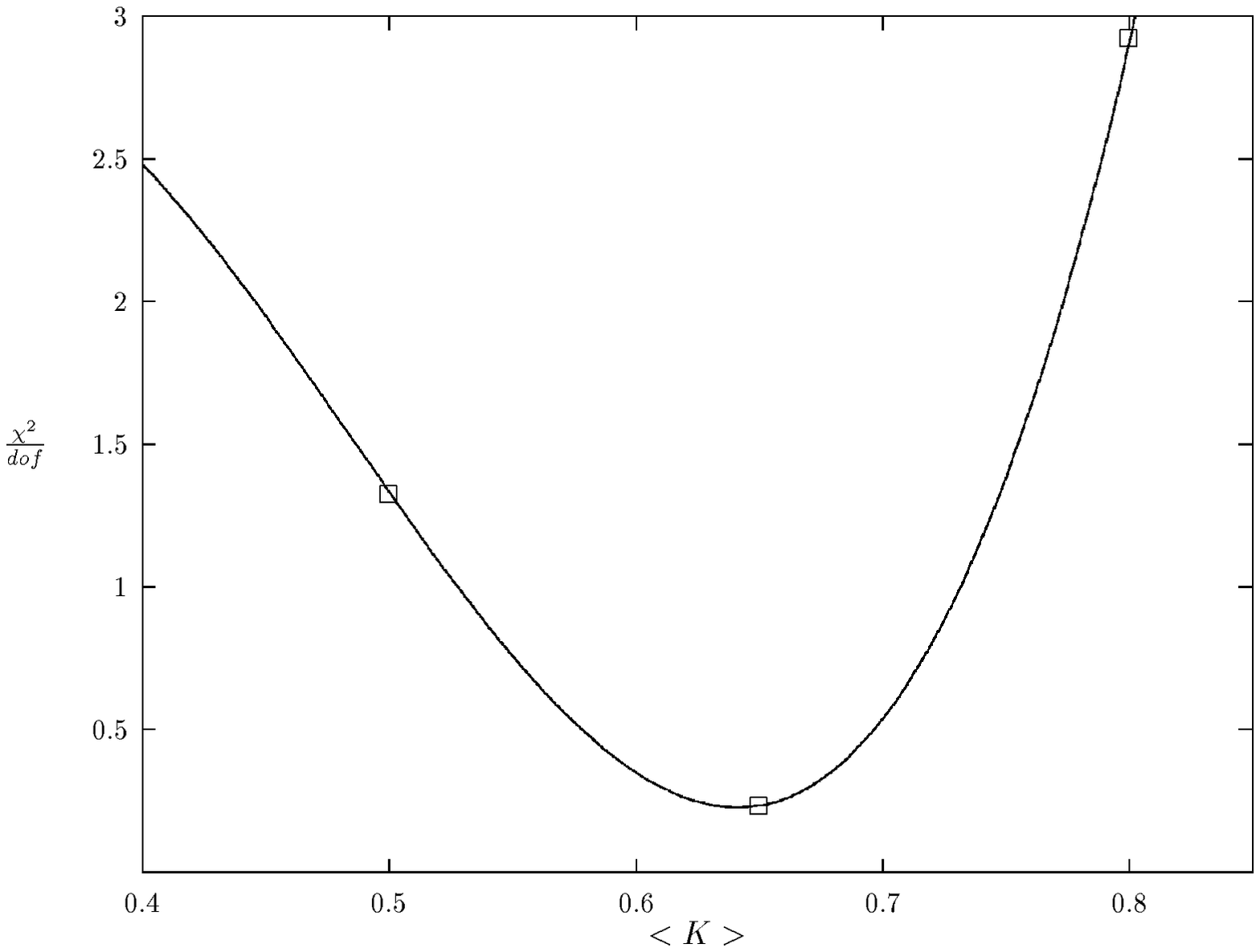}
\end{picture}
\end{figure}
\newpage
\centerline{Figure 5}
\begin{figure}[h]
\setlength{\unitlength}{1cm}
\begin{picture}(15.,10.)
\includegraphics{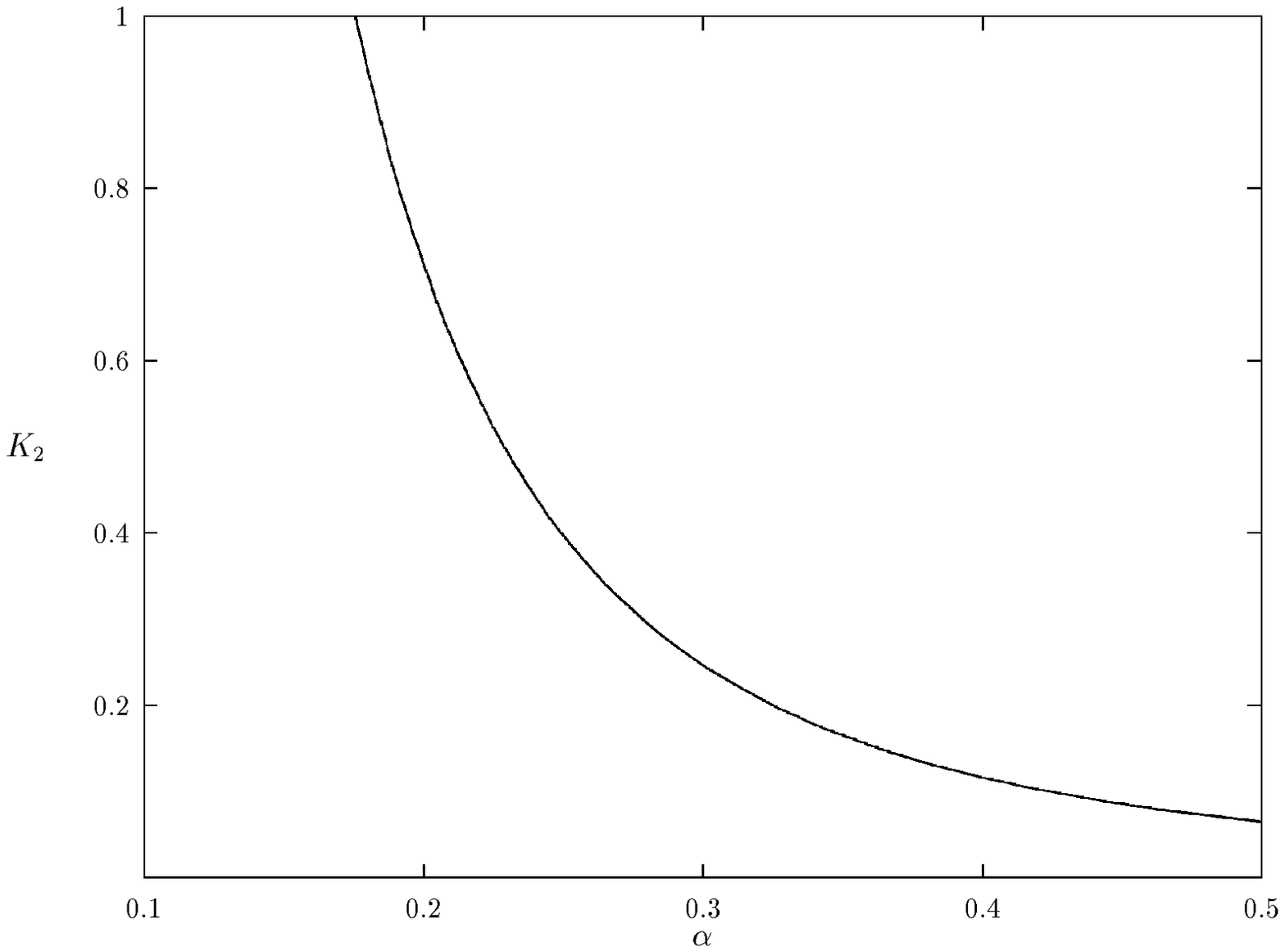}
\end{picture}
\end{figure}


\begin{thebibliography}{99}
\bibitem{FOOT1} This fact means that experimentally selected
                inelasticity will be distorted and equal to
                the spectrum weighted one:
                $$\langle K_{exp}\rangle = \langle K^{\gamma - 1}\rangle
                ^{\frac{1}{1 - \gamma}},$$
                where $K$ is inelasticity from eqs.(\ref{eq:epsilons},
                \ref{eq:etas}) above and $\langle \dots \rangle$
                means averaging over the energy spectrum $N(E_0)$.
\bibitem{FOOT2} At the $\sim 600$ g/cm$^2$ of depth of atmosphere
                where Pamir experiments were performed all heavier
               (i.e., nuclear) components of incoming flux are
               already dissociated into nucleons in the first tens of
               g/cm$^2$ of atmosphere. This all-nucleon flux is the
               enriched with mesonic (for all practical purposes -
               pionic) component produced in subsequent interactions
               in the atmosphere. The relative percentage of nucleons
               and pions can be fixed from other cosmic ray
               experiments \cite{W1}.
\bibitem{FOOT3} Notice that in reality the transverse dimension of
                regions where energies $E_1$ and $E_2$ are deposited
                are both very small, of the order of $100~\mu$m, and
                that particles in cascade are not separated
                experimentally.
\bibitem{FOOT4} Although our results, as well as those of \cite{B},
                are based on rather low statistics, they seem to be
                the only results for inelasticity available for
                nuclear targets for foreseable future (at least for
                the energies above $E_{LAB} > 100$ TeV). The
                remarkable feature of both EC's allows for the most
                direct etimation of $K$ done so far.
\end{thebibliography}
\end{document}